# Integrated Arbitrary Filter with Spiral Gratings: Design and Characterization

Yi-Wen Hu, Shengjie Xie, Jiahao Zhan, Yang Zhang, Sylvain Veilleux, and Mario Dagenais, *Fellow, IEEE*

*Abstract*—We report the design and characterization of a high performance integrated arbitrary filter from 1450 nm to 1640 nm. The filter's target spectrum is chosen to suppress the night-sky OH emission lines, which is critical for ground-based astronomical telescopes. This type of filter is featured by its large spectral range, high rejection ratio and narrow notch width. Traditionally it is only successfully accomplished with fiber Bragg gratings. The technique we demonstrate here is proven to be very efficient for on-chip platforms, which can bring many benefits for device footprint, performance and cost. For the design part, two inverse scattering algorithms are compared, the frequency domain discrete layer-peeling (*f*-DLP) and the time domain discrete layer-peeling (*t*-DLP). *f*-DLP is found to be superior for the grating reconstruction in terms of accuracy and robustness. A method is proposed to resolve the non-uniformity issue caused by the non-zero layer size in the DLP algorithm. The designed 55-notch filter is 50-mm-long and implemented on a compact $Si_3N_4/SiO_2$ spiral waveguide with a total length of 63 mm. Experimentally, we demonstrate that the device has an insertion loss as low as 2.5 dB, and that the waveguide propagation loss is as low as 0.10 dB/cm. We are also able to achieve uniform notch depths and 3-dB widths of about 28 dB and 0.22 nm, respectively.

## I. INTRODUCTION

HIGHLY selective optical filters are fundamental building blocks for applications in many areas such as optical communications [1], [2], signal processing [3], [4], optical sensing [5], quantum information [6], [7] and astronomy [8], [9]. Over the past decades, a lot of works in these areas have been done with fiber-optic technology, most prominently the fiber Bragg grating (FBG). With the development of fiber grating inscription methods, such as the phase mask and the direct writing technique, now people can fabricate ultra-long, complicated and accurate FBGs [10]–[12].

In recent years there is an ever-increasing demand for on-chip optical interconnection which can realize most possible functions in a small footprint. Within these objectives, high performance on-chip arbitrary photonic filters are an important aspect [13]–[15]. As an example, for ground-based astronomical telescopes, several hundred lines are needed with precise positions and depths to suppress the night-sky OH emissions from the Earth's atmosphere. This type of filter is featured by its large spectral range, high rejection ratio and narrow notch width, which up to now is best achieved with complex Bragg gratings. Several studies have been carried out using FBGs to achieve this type of filter [9], [12], [16]–[18]. With the rapid advancements of various chip-based fabrication techniques, many exciting works have been done with integrated Bragg gratings [19]–[21]. However, the devices which can be used as this type of arbitrary filter are still rare. Our group is the first to demonstrate an arbitrary filter with complex waveguide Bragg grating (CWBG) on a $Si_3N_4/SiO_2$ CMOS compatible platform [14]. Nonetheless, there remain quite a few issues with the prototype device such as irregular notch depths, limited spectral range, and high insertion loss. These issues will severely limit our ability to put the integrated CWBG into practical use. The focus in this work is to resolve these issues and obtain a much-improved filter response.

The essence of the CWBG design is an inverse scattering (IS) problem. Among various IS algorithms for grating reconstruction, the layer-peeling (LP) algorithm is one of the most popular because of its accuracy, robustness, and speed [22]–[25]. Differentiated by the numerical implementation method, LP can be classified as discrete layer-peeling (DLP) or continuous layer-peeling (CLP). DLP is more adopted in recent works because of its high efficiency and self-consistency [24]. Furthermore, like many other electromagnetic field problems, DLP can be implemented in frequency domain (*f*-DLP) or time domain (*t*-DLP) [26]. The performance of both will later be discussed within our application scenario.

The remainder of this paper is organized as follows. In Section II, we address the theoretical issues and optimization of the DLP algorithm. Next in Section III, we discuss the details of the spiral waveguide design. Section IV and Section V mostly include the experimental results from device fabrication to characterization. A brief conclusion is made in Section VI.

Manuscript submitted February 21, 2020; date of current version April 24, 2020. This work was supported by the National Science Foundation under grant 1711377 and National Aeronautics and Space Administration under grant 16-APRA 16-0064.

Y.-W. Hu, S. Xie, J. Zhan, Y. Zhang and M. Dagenais are with the Department of Electrical and Computer Engineering, University of Maryland, College Park, MD 20742 USA (e-mail: yiwenhu@umd.edu; dage@umd.edu).
S. Veilleux is with the Department of Astronomy, University of Maryland, College Park, MD 20742 (email: veilleux@astro.umd.edu)

## II. Optimization on Discrete Layer-Peeling

First, we will provide a concise description of the DLP algorithm. Detailed discussions can be found in earlier works [14], [24]. DLP starts with a target spectrum which can be almost arbitrarily chosen. Then it divides a grating into $N$ layers (also referred as pieces in this paper), with all layers having the same layer size (piece length) $\Delta$. The whole grating has a length $N \times \Delta$ and can be reconstructed from the target spectrum with $N$ iterations. In each iteration, the foremost layer of the remaining grating can be determined from the in-situ field by the law of causality. The next iteration is executed after the field propagate by a distance $\Delta$ and the previous layer is peeled off. The algorithm completes when the last layer is reached.

Once DLP finishes, the discrete layer adding (DLA) algorithm can be conveniently used to validate the reconstructed grating. DLA is a typical direct scattering (DS) algorithm, which is also important for filter design. However, we should mention that DLA is not a proper DS algorithm, which will be explained later in this section. A preferred DS algorithm is highly consistent with the experiment and can be used to value the performances of different IS models. In this section we use two DS models to validate our DLP. The first one is the piecewise transfer matrix $T^{\mathrm{CMT}}$ derived from the coupled-mode theory (CMT), with its pieces being the same as with DLP. It is the most intuitive validation model as LP is inherently based on CMT. The other one is the ABCD matrix model, in which we sample the sinusoidal-like grating into short rectangular segments, with the segment length much shorter than the grating period. The corresponding transfer matrix is derived by imposing the electromagnetic continuity conditions at segment boundaries [27]. Although this ABCD model is best suited for quasi-periodic structures, we will see later that it performs well for simulating our complex aperiodic gratings.

### A. Frequency domain and time domain DLP

To compare $f$-DLP and $t$-DLP, we select 55 lines and 36 lines near 1550 nm from the night-sky OH emission lines [28]. The target spectra are shown in the top two panels of Fig. 1. They are determined by the following equations [29],

$$r_n = \frac{\sqrt{R_n}}{\cosh\{\operatorname{arcosh}(\sqrt{2}) \times [(\lambda - \lambda_n)/(w_n/2)]^2\}} \quad (1)$$

$$r_{\mathrm{sum}} = \left[1 - \prod_{n=1}^{N}(1 - |r_n|^2)\right]^{1/2} \quad (2)$$

Here $r_n$ is the reflection amplitude for the $n$-th notch with $n = 1, 2, 3\ldots N$, and $r_{\mathrm{sum}}$ is the total reflection amplitude. $R_n$ is set so that the notch depths are 15 dB or 30 dB. The full width at half maximum (FWHM) $w_n$ is set as 0.2 nm. $r_{\mathrm{sum}}$ is the total reflection amplitude. The layer size $\Delta$ is 4 μm in Fig. 1(a), and is 8 μm in Fig. 1(b). Note that Fig. 1(b) has a narrower spectral range because the DLP algorithm's bandwidth $\delta_{\mathrm{BW}} = \pi/\Delta$.

We apply $t$-DLP and $f$-DLP to the two target spectra respectively and validate the four reconstructed gratings with the CMT model. As shown in Fig. 1, for notches with 15 dB depth, $t$-DLP and $f$-DLP both work well. For 30 dB notches, however, $t$-DLP generates some irregularities at the longer wavelength side. It shows that $f$-DLP gives more robust results, although $t$-DLP is faster in speed [25]. $f$-DLP is superior over $t$-DLP if precise control over individual notch lineshape is a top priority.

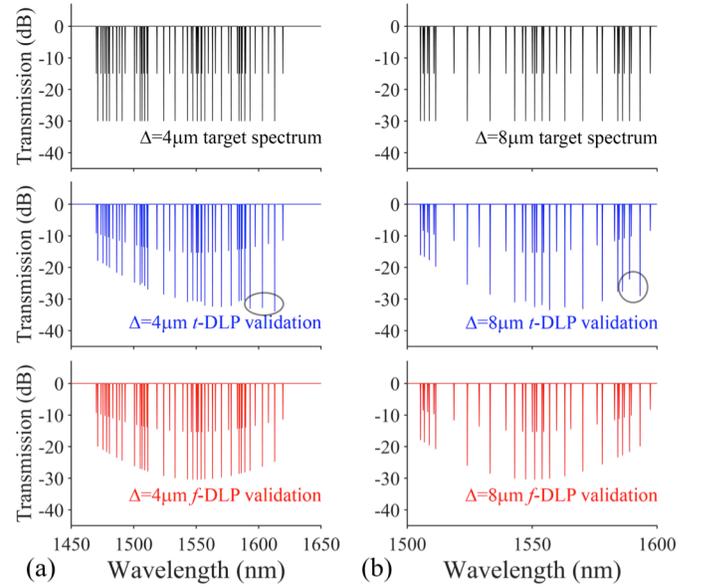

Fig. 1. Comparison of $t$-DLP and $f$-DLP for an arbitrary filter design with 15dB/30dB alternating notch depths. (a) DLP layer size is 4 μm. (b) DLP layer size is 8 μm. Grey ovals indicate the irregular reconstructed notches in the $t$-DLP algorithm. Both $t$-DLP and $f$-DLP have the non-uniformity issue.

### B. Non-uniformity issue

Another obvious issue in the validation plots is the non-uniform notch depths. In both $t$-DLP and $f$-DLP, the notch depths only match well with the target near the bandwidth center, and become shallower elsewhere, especially near the bandwidth boundaries. As we have mentioned, a non-zero $\Delta$ corresponds to a limited $\delta_{\mathrm{BW}}$. Currently, a $\Delta$ of 4 μm is about the minimum layer size we can adopt, which corresponds to a wavelength span of about 190 nm. For most optical filter applications, it is much larger than the span of the target spectrum, so this issue will not be a huge concern. However, for our OH suppression filter it can cause serious performance impairment.

We find that this non-uniformity issue is caused by the conflict between the discrete localized reflector model and the actual continuous grating structure. In the rest of this section, we will take $f$-DLP with $\Delta = 4$ μm as the example to elaborate. The target spectrum $r_{\mathrm{sum}}$ is sampled by $M$ points to form a sampled target spectrum $r_s$. $r_s$ and its corresponding pulse response $h$ are a discrete Fourier transform (DFT) pair,

$$r_s(j) = \sum_{k=0}^{M-1} h(k)e^{i2\pi \cdot \delta_j \cdot k 2\Delta/2\pi} \quad (3)$$

$$h(k) = \frac{1}{M}\sum_{j=0}^{M-1} r_s(j)e^{-i2\pi \cdot \delta_j \cdot k 2\Delta/2\pi} \quad (4)$$

where $\delta_j, j = 1, 2\ldots M$ are equally spaced within $[-\pi/2\Delta, \pi/2\Delta] \sim [-4\times10^5, 4\times10^5]$, the detuning range of wavenumber $\delta$. $r_s$ should be perfectly reconstructed if $M$ is large enough and we have the right pulse response $h$. To connect $h$ with an actual grating structure, CMT transfer matrix $T^{\text{CMT}}$ is used to analyze the field propagation. We denote $q_n$ as the coupling coefficient of the $n$-th grating piece. For the target spectrum in this work, $|q_n|$ is relatively small, always less than $10^4$ m$^{-1}$. Therefore $\delta \gg |q_n|$ for most of $\delta$. The transfer matrix of the $n$-th grating piece $T_n^{\text{CMT}}$ can then be simplified as

$$T_n^{\text{CMT}} \cong \begin{bmatrix} e^{i\delta\Delta} & q_n \sin(\delta\Delta)/\delta \\ q_n^* \sin(\delta\Delta)/\delta & e^{-i\delta\Delta} \end{bmatrix} \quad (5)$$

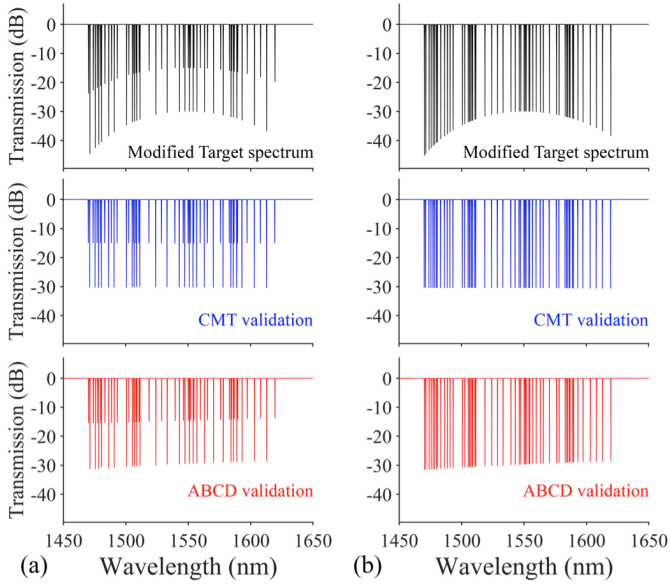

Fig. 2. $f$-DLP modified by a sinc function for a target spectrum with (a) alternating 15dB/30dB notch depths and (b) uniform 30dB notch depths. Validation plots show that the modified $f$-DLP does not have the non-uniformity issue.

In the DLP model, the grating is assumed to consist of discrete localized reflectors. The reflector spacing corresponds to the pulse response interval. As a result, the more accurate transfer matrix should not be $T_n^{\text{CMT}}$, but the product of a pure propagation matrix and a pure reflector matrix $T^\Delta \times T^\rho$, with

$$T^\Delta = \begin{bmatrix} e^{i\delta\Delta} & 0 \\ 0 & e^{-i\delta\Delta} \end{bmatrix} \quad (6)$$

$$T_n^\rho = \frac{1}{\sqrt{1-|\rho_n|^2}} \begin{bmatrix} 1 & \rho_n^* \\ \rho_n & 1 \end{bmatrix} \quad (7)$$

$$T^\Delta \times T_n^\rho \overset{|q_n\Delta| \ll 1}{\cong} \begin{bmatrix} e^{i\delta\Delta} & q_n \cdot e^{i\delta\Delta} \\ q_n^* \cdot e^{-i\delta\Delta} & e^{-i\delta\Delta} \end{bmatrix} \quad (8)$$

where $\rho_n = \tanh(|q_n\Delta|)q_n^*/|q_n|$ is the complex reflection coefficient. DLP is based on (3), (4), (6), (7), and we can see that under the discrete localized reflector assumption DLP is a very self-consistent model. In other words, the spectrum calculated by the discrete layer adding algorithm (DLA) will usually match well with the target in DLP (this is noticed but not explained in [24]). However, for an actual grating, the grating structure is continuous and $T_n^{\text{CMT}}$ should be the transfer matrix to use for field propagation. This is the reason why non-uniformity occurs.

Our proposed method to fix this issue is explained below. When $\delta \gg |q_n|$, if we use

$$q_n^{\text{adjust}} = q_n e^{i\delta\Delta} \cdot \frac{\delta\Delta}{\sin(\delta\Delta)} \quad (9)$$

$T^\Delta \times T_n^\rho$ becomes equal to $T_n^{\text{CMT}}$ in (5). When $\delta \gg |q_n|$ is not satisfied, (5) will not hold. We choose the layer size $\Delta$ small enough such that $\delta\Delta \ll 1$ for these smaller $\delta$, thus $q_n^{\text{adjust}} \cong q_n$. In other words, we are not changing the original DLP near the bandwidth center, which is good because $T^\Delta \times T^\rho$ strictly equals to $T^{\text{CMT}}$ at $\delta = 0$. As a result, the adjusted grating profile should be able to solve the non-uniformity issue. Note that the amplitude adjustment factor in (9) is the reciprocal of a sinc function. However, for such a grating structure, it is very difficult to have a wavelength dependent coupling coefficient determined by (9). A feasible approach is to modify the target spectrum such that (9) "seems" to be satisfied. Specifically, with CMT we can calculate an equivalent $\tilde{q}$ according to the target notch depth and width. Next we calculate $\tilde{q}^{\text{adjust}}$ from $\tilde{q}$ with (9) and modify each notch's depth and width in the original target spectrum according to $\tilde{q}^{\text{adjust}}$. Two target spectra such modified are shown in the top of Fig. 2. We apply $f$-DLP to them, and the CMT validation plots show that the non-uniformity indeed disappears. The ABCD validation plots below show a very similar transmission, although the notch depths are not as uniform as the CMT's (notch depths are slightly shallower at the longer wavelengths).

Above all, the non-uniformity issue of DLP vanishes after we resolve the conflict between the discrete model and the actual continuous grating segments. Since CLP uses $T^{\text{CMT}}$ for a continuous coupling process, we may think that it will not have this issue. This is not true by our study, with similar non-uniformity phenomenon also found in CLP. In fact, the fundamental problem lies in the discretization process itself, which is inevitable for any numerical implementation of LP. In either DLP or CLP, the discretization process needs to assume that the grating is uniform over a certain distance, which does not fully obey causality unless $\Delta \rightarrow 0$.

### III. Spiral Complex Grating Design

#### A. Spiral curve geometry

For the 30 dB/0.2 nm filter design we discussed in this paper, gratings of several cm or even longer are typically required. By mapping the complex grating onto a spiral waveguide structure, we can significantly reduce the footprint for better fabrication uniformity (e.g. less stitching error) and efficient chip integration. Meanwhile, we also need to carefully design the spiral structure so that it brings minimal negative impact when compared with a straight structure. Furthermore, different from a simple spiral waveguide, for the spiral complex grating design, we must consider the geometry of each grating piece and combine them together.

The Archimedean spiral is used to define the spiral curve. In a polar coordinate system, it can be written as

$$\rho = \frac{\Delta r}{2\pi}\theta \quad (10)$$

where $\Delta r$ determines the spiral waveguide spacing. The arc length of this curve when angular coordinate goes from 0 to $\theta$ is

$$s(\theta) = \frac{\Delta r}{4\pi}\left[\theta\sqrt{1+\theta^2} + \ln\left(\theta + \sqrt{1+\theta^2}\right)\right] \quad (11)$$

The slope of the curve at $\theta$ is

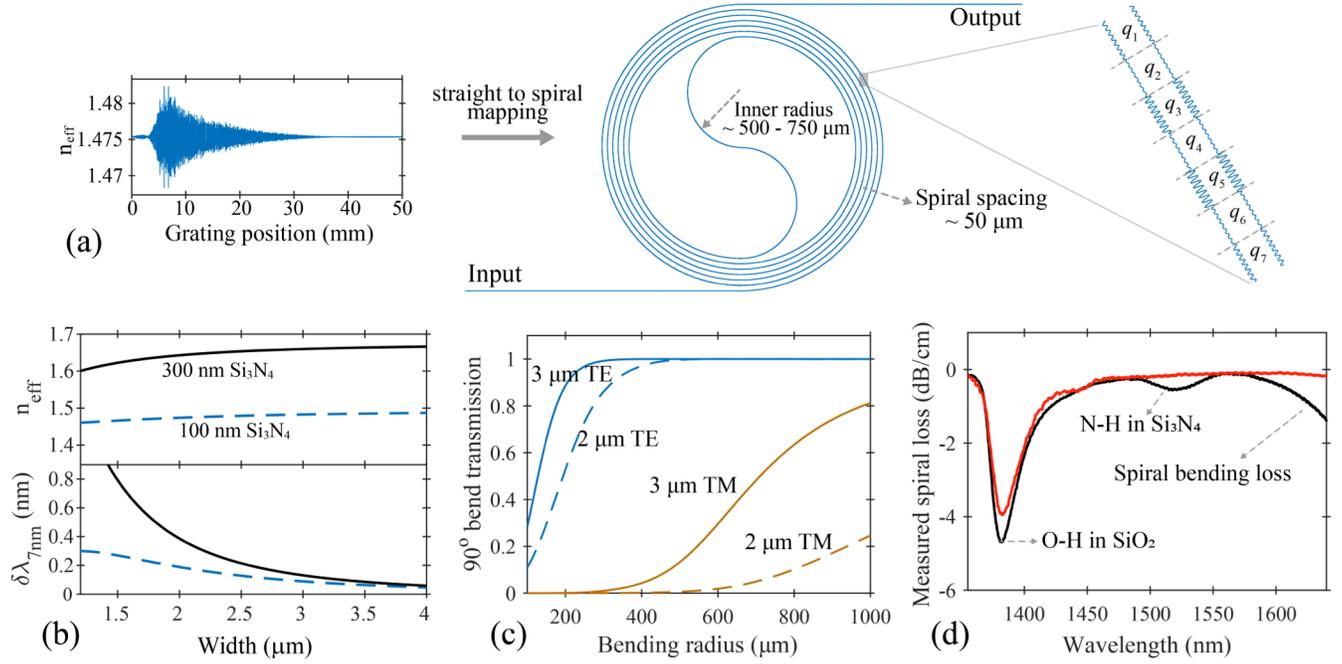

Fig. 3. (a) Illustration of a spiral complex grating. The left panel shows the index profile of the 50 mm grating. The right panel shows the whole grating, assembled by a series of layers with different coupling coefficient $q_i$. (b) Mode index study for waveguides with 100 nm and 300 nm thicknesses. (c) Simulated bending loss in TE and TM modes at 1550 nm for 3 μm x 100 nm and 2 μm x 100 nm waveguides. (d) Measured spiral losses for two spiral waveguides with 500 μm inner radius (black) and 750 μm inner radius (red). The 750 μm inner radius sample is annealed in 1150 .

$$m = \frac{\theta + \tan\theta}{1 - \theta\tan\theta} \quad (12)$$

The spiral curve geometry is determined from (10) and (11). Specifically, we design a counter-clockwise spiral and a clockwise spiral in the layout. The two spirals are connected with two half circles with radius $\rho_0$. $\rho_0$ is chosen to be larger than the critical bending radius to avoid noticeable bending loss. Note that the grating structure is mapped onto the full curve in in Fig. 3(a). This full curve includes two spirals, two half circles, and two short straight regions at the input/output ends. For any required grating length, we adjust $\rho_0$, $\Delta r$ and the length of the straight region so that a satisfactory curve is obtained.

Once we have the curve geometry, we discretize the continuous curve into short straight pieces. Each piece is a layer in DLP, and contains a grating structure with a uniform coupling coefficient $q_i$. For the design we use here, the length of each piece is $\Delta = 4$ μm and the 50-mm-long grating corresponds to $N = 12500$ pieces. They are all positioned according to the starting phase $\theta_i$ and the slope $m_i$, with $i = 1, 2, 3 \ldots 12500$. $m_i$ is calculated from $\theta_i$ by (12), and $\theta_i$ is determined by $\theta_i = \theta_{i-1} + \varphi_i$, where $\varphi_i$ is the angular coordinate change brought by the $i$-th grating piece.

The grating profile calculated from $f$-DLP based on the target spectrum in Fig. 2(b) is shown on the left of Fig. 3(a). We map this 50-mm-long grating to a spiral waveguide layout in FIMMPROP with MATLAB scripting. As is illustrated, it consists of two straight regions, two Archimedean spirals and two connecting half circles. It is worth noting that this in-plane straight to curve mapping procedure is not a distance preserving transformation, and will inevitably introduce geometric errors. The inner side of the waveguide grating will suffer a length cut, while the outer side of grating will suffer a length extension. However, our experimental results in Section V show that this error is negligible for our application.

### B. Mode index's sensitivity on waveguide width

As illustrated in Fig. 3(a), the 50 mm long grating device is curled to a size of 1 mm – 1.5 mm, much smaller than its fiber counterpart which typically can't be bent. The footprint is mainly limited by the critical bending radius and can be further reduced if we use a thicker $Si_3N_4$ waveguide. This can be proved from the top part of Fig. 3(b), where we simulate the mode indices for waveguides with 100 nm and 300 nm $Si_3N_4$ thicknesses. Note that a higher mode index implies a stronger mode confinement.

On the other hand, there is another trade-off factor to consider when choosing the $Si_3N_4$ thickness. For the thicker nitride waveguide, its mode index is higher, but is also more sensitive to width variations. This is undesired for narrow filter design because for any lithography system, the minimum feature $\delta w$ it can provide is limited. Here we assume that the single side width variation $\delta w/2 = 7$ nm, which is the line width limit for our Elionix electron-beam lithography (EBL) system, and denote $n_1$, $n_2$ as the mode index for width $w$ and $w + \delta w$, respectively. For a sinusoidal Bragg grating, the notch width limit $\delta\lambda$ can be estimated as $\delta\lambda = \lambda(n_2-n_1)/(n_2+n_1)$. The simulation result is plotted in the bottom figure of Fig. 3(b). If we want $\delta\lambda < 0.2$ nm, 300 nm thick waveguide needs to have a width over 2.6 μm, and 100 nm thick waveguide needs to have a width over 2.0 μm. Considering that the single mode width limit is ~1.3 μm for 300 nm and ~3 μm for 100 nm, the 100 nm $Si_3N_4$ is chosen for the final design.

### C. Critical bending radius

Compared with a conventional straight CWBG design, the spiral design could potentially bring noticeable bending losses. In Fig. 3(c), we simulate the transmission of TE and TM modes propagating through a 90-degree bend waveguide section in FIMMWAVE. Two $Si_3N_4$ core dimensions are studied: 2 μm (W) × 100 nm (H) and 3 μm × 100 nm. We will focus on the TE mode because of its stronger mode confinement and much lower bending loss [27]. In the simulation, the critical bending radius for TE mode is about 500 μm for the 2 μm wide waveguide and 300 μm for the 3 μm wide waveguide. In the experiment, however, the critical radius will be larger due to light scattering from surface/boundary roughness.

In Fig. 3(d), we measured the spiral loss by subtracting two 50 mm spiral uniform waveguides with a reference straight waveguide. The black curve is the loss for a 3 μm wide spiral waveguide with 500 μm inner radius; the red curve is the loss for an annealed 3 μm spiral waveguide with 750 μm inner radius. Three dominant sources of loss are illustrated. The left one is caused by O-H bonding in PECVD $SiO_2$; the middle one is caused by N-H bonding in LPCVD $Si_3N_4$, which can be removed by annealing; the right one is caused by the bending spiral structure. As can be seen, when increasing the inner radius from 500 μm (black curve) to 750 μm (red curve), the transmission drop caused by bending loss at longer wavelengths disappears.

## IV. Fabrication and Loss Measurement

The fabrication process is similar to our previous work [27]. We use a 100 kV Elionix ELS-G100 e-beam lithography system with negative resist Ma-N 2403 to pattern the structure. For the PECVD $SiO_2$ top cladding, we switch from the $SiH_4$ recipe ($SiH_4$, $NH_3$ and $N_2$) to TEOS recipe ($Si(OC_2H_5)_4$ and $O_2$). O-H bond absorption near 1.4 μm will still exist. However, with the absence of nitrogen content, TEOS recipe can eliminate N-H bond absorption in PECVD $SiO_2$ film near 1.5 μm. Since we mainly focus on the spectral range between 1450 nm and 1640 nm, the only dominant loss left is the N-H absorption in LPCVD $Si_3N_4$ film indicated on the black curve of Fig. 3(d). As we have mentioned in Section III(C), an effective solution is high temperature (1150  ) annealing [27]. The loss curve after annealing is shown in red, and clearly shows that N-H absorption is gone. Note that unlike many other works, we could not perform the annealing after the patterning step because the sidewall variation structure can be altered when experiencing such a high temperature.

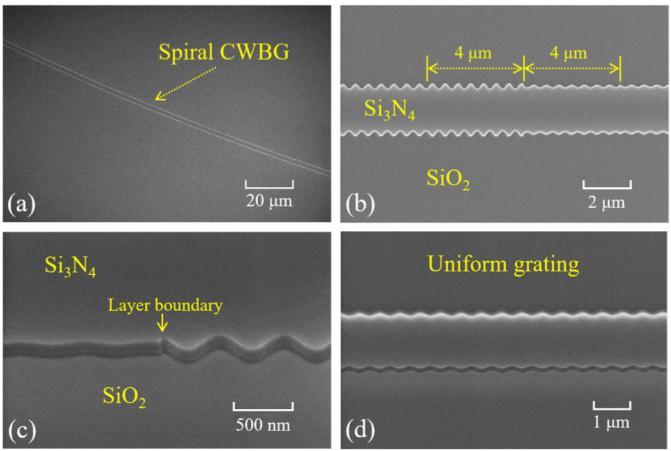

Fig. 4. SEM of two fabricated devices. (a) A segment of the spiral CWBG. (b) Zoomed-in view of (a). (c) Tilted and further zoomed-in view of (a). (d) Tilted view of another uniform grating used as an FPBG mirror.

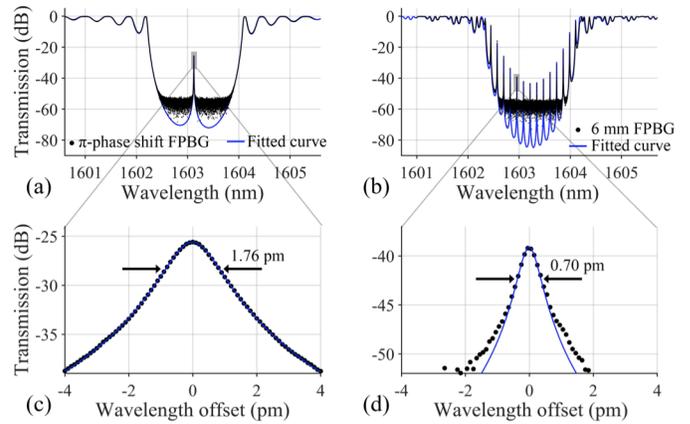

Fig. 5. Waveguide and grating loss characterization based on FPBG cavities. The grating loss is extracted as 0.26 dB from (a) π-phase shift FPBG. The waveguide loss is extracted as 0.10 dB from (b) 6mm FPBG. (c),(d) are zoomed-in views of (a),(b).

Limited by field of view of SEM, we could only capture a small portion of the curved grating as shown in Fig. 4(a). A slight bend can still be observed. Enlarged view in Fig. 4(b,c,d) shows that the grating sidewall corrugation is well defined, which is important for low-loss applications. SEM in Fig. 4 shows that the grating sidewall corrugation is well defined, which is important for low-loss applications. Compared with our previous work on complex gratings [14], [30], we use a continuous sinusoidal structure instead of sampled discrete rectangular pieces. This optimization makes our grating design more consistent with the DLP model. It can provide much smoother grating boundary, reduced on-chip loss, and improved spectrum accuracy. In Fig. 4(b) and 4(c), we can see the grating consists of 4 μm layers. Each layer has a coupling coefficient $q_i$ determined from $f$-DLP.

In Fig. 3(d), we have roughly measured the spiral loss with the reference waveguide approach. To characterize the waveguide and grating losses in a more accurate way, we use the Fabry-Perot Bragg grating cavities (FPBGs) method as demonstrated in [27]. The cavity mirrors are 2-2.4 μm uniform Bragg gratings with 3000 periods on each side. A part of the grating mirror is shown in Fig. 4(d). The loaded quality factor is 0.94 million for the π-phase shift FPBG and 2.3 million for the 6 mm FPBG. By fitting the experimental curve to our grating model, we determine that the grating mirror loss is 0.26 dB/cm and the uniform waveguide loss is 0.10 dB/cm.

## V. Spiral Grating Characterization

We have fabricated a 55-dip CWBG spiral and a reference spiral on a single chip and the measured insertion loss is shown in Fig. 6(a). The insertion loss is defined as the difference of the measured transmission with and without the chip coupled. This CWBG is designed with the modified *f*-DLP with grating length 50 mm and total length 63 mm. Waveguide core dimension is 3 µm × 100 nm. This sample is also annealed at 1150 before EBL to get a flat transmission. The reference waveguide has an identical spiral structure as the spiral CWBG, but does not have any sidewall corrugations. For the reference spiral, the insertion loss (from input fiber to output fiber) is 1.6 dB – 2.8 dB across the full spectrum. Such a high throughput is achieved with a Nufern UHNA3 fiber, which can realize a fiber-to-chip coupling efficiency > 90% per cleaving facet [31]. For the spiral CWBG, the lowest insertion loss is 2.5 dB near 1580 nm but increases to ∼ 11 dB near 1450 nm. By subtracting the lowest losses of the two spirals and dividing it by the spiral length, we obtain that the average CWBG loss is 0.18 dB/cm higher than for a uniform waveguide (∼ 0.10 dB/cm from Fig. 5). Furthermore, this sample does not exhibit any obvious non-uniformity. Most of the notch depths are around 28 dB, and all are between 25 dB and 30 dB.

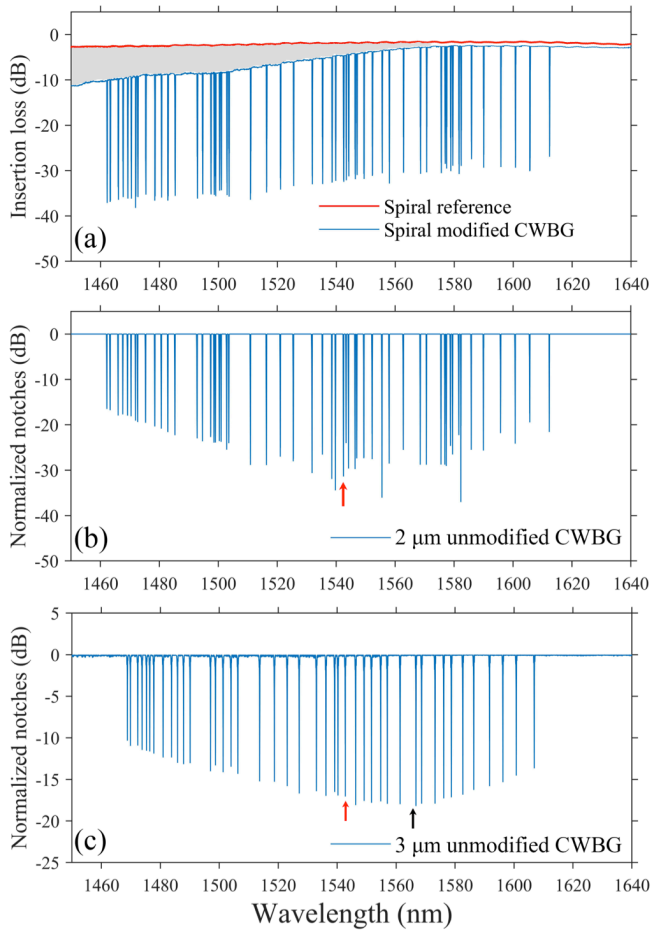

Fig. 6. (a) Experimentally measured transmission of a reference spiral and a CWBG spiral with 3 µm average width. (b) 2 µm average width CWBG with 55-notch design showing that the deepest notches lie near the bandwidth center. (c) 3 µm average width CWBG with 43-notch design showing that the deepest notches are shifted to the right from the bandwidth center. Note that the notches in (b) and (c) are all normalized to 0 dB to compare notch depths.

As a comparison to Fig. 6(a), we design and fabricate two spiral filters with the unmodified *f*-DLP with 2 µm and 3 µm wide waveguide, i.e without the sinc function adjustment in (9). The results are shown in Fig. 6(b, c). Note that in these two plots all notches are normalized to 0 dB so that individual notch depth can be easily compared. Both curves indeed have the non-uniformity issue. The notch bottom contour is very similar with CMT validation plots in Fig. 1. Compared with Fig. 6(a), it clearly shows that the non-uniformity issue can be well resolved with the sinc function adjustment. These graphs also confirm that our CMT and ABCD validation models match the actual experiment pretty well.

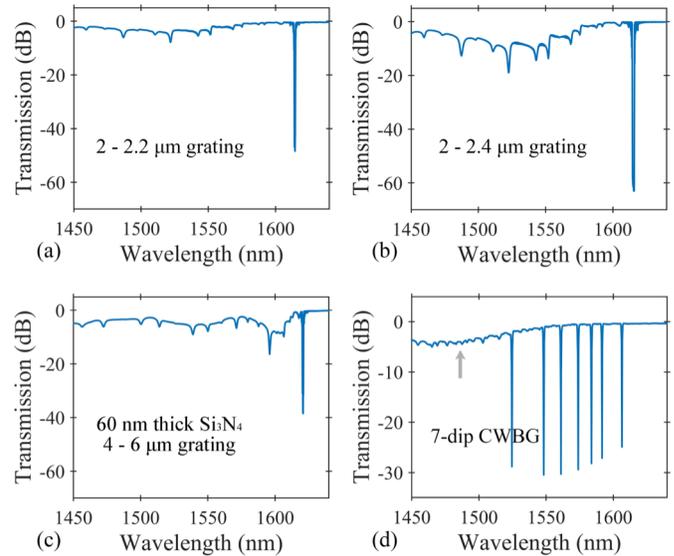

Fig. 7. The study on cladding-mode coupling. Transmission of (a) 100 nm thick $Si_3N_4$ with 2 − 2.2 µm wide single-notch grating; (b) 100 nm thick $Si_3N_4$ with 2 − 2.4 µm wide single-notch grating; (c) 60 nm thick $Si_3N_4$ with 4 − 6 µm wide single-notch grating; (d) 7-notch CWBG with FWHM ∼ 0.8 nm.

### A. Effect from cladding mode coupling

As shown in the grey region of Fig. 6(a), there is an obvious intensity drop for the spiral CWBG transmission at shorter wavelengths. This drop pattern is mostly caused by the grating structures, more specifically the coupling between the forward guided mode and the backward cladding mode in CWBG. It is a quite common phenomenon for Bragg grating devices. It was first observed and studied in FBGs and then in SOI gratings [32]–[34]. A detailed discussion of cladding mode coupling on the $Si_3N_4/SiO_2$ platform can be found in our recent paper [35].

This effect will typically result in some distinct dips on the higher-frequency side of the main stopband, as seen in Fig. 7(a, b, c), which are experimentally measured transmission for specified simple Bragg gratings. We notice that the drop pattern is more obvious for stronger gratings, and it will move closer to the main stopband when $Si_3N_4$ is thinner. For a CWBG, which can be regarded as a combination of some simple Bragg gratings, many distinct dips could appear and overlap, as seen in Fig. 7(d). If a CWBG contains too many dips, we could see a continuous drop pattern as seen in Fig. 6(a).

Since cladding mode coupling could significantly decrease the complex filter throughput, for our application it is undesired and should be avoided. For CWBG in this work, the most promising method might be adopting a weaker grating design. Note that even if the target spectrum is unchanged, we can still decrease the maximum grating strength by introducing different group delay terms for different notches [18]. It will stretch the pulse response $h$ in time and lead to a longer grating in space, which needs to be considered as a trade-off factor in practice. Further discussions on how to avoid this drop pattern can be found in [10], [35].

### B. Effect from waveguide dispersion

Dispersion is always an important factor for multi-notch filter design. In our grating structure, the waveguide mode index $n_{eff}(\lambda, d)$ is a function of both wavelength $\lambda$ and waveguide width $d$. To precisely align each notch to the target spectrum, one usually needs to take care of the first-order partial derivative $\partial n_{eff}/\partial \lambda$. If it is not accurately determined, the whole spectrum will look like expanded or compressed along the horizontal axis. Moreover, there is another effect also caused by waveguide dispersion, which is related with the second-order partial derivative $\partial^2 n_{eff}/\partial \lambda \partial d$. The fact that this derivative is not zero implies that, for a given width variation, the coupling coefficient is different for light with different wavelengths. In Table 1, we simulate several simple Bragg gratings designed at three different wavelengths with the rigorous coupled-mode theory (RCMT) algorithm in FIMMPROP. For 2-2.1 μm and 2.4-2.5 μm width variations, the notch depths differ by 1.1 dB and 1.3 dB. However, for 3-3.1 μm width variations, the notch depths variation can reach 3.5 dB.

TABLE I
GRATING NOTCH DEPTH DIFFERENCE CAUSED BY DISPERSION

|  | 2μm - 2.1μm[a] | 2.4μm - 2.5μm | 3μm - 3.1μm |
|---|---|---|---|
| 1450 nm | 29.5 dB | 29.3 dB | 28.8 dB |
| 1550 nm | 29.4 dB | 30.4 dB | 31.0 dB |
| 1640 nm | 28.4 dB | 30.6 dB | 32.3 dB |

[a] denotes the widths of narrow – wide part of a simple Bragg grating.

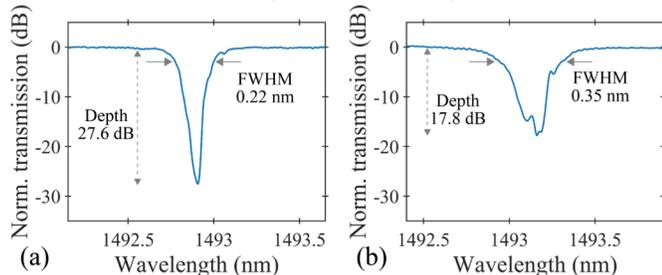

Fig. 8. Normalized transmission of a well-defined notch and a disturbed notch. (a) The zoomed-in view of a single notch in Fig. 6(a). (b) The transmission of a 2 μm width spiral CWBG with fabrication errors.

This effect is observed in our experiments. Recall that without sinc function adjustment, the deepest notch will always lie near the bandwidth center due to the non-uniformity issue. In Fig. 6(b, c), we mark the bandwidth center position with the red arrow. For 2 μm unmodified CWBG, the red arrow indeed corresponds to the deepest notch position. For 3 μm unmodified CWBG, however, the deepest notch position (black arrow) differ from the bandwidth center (red arrow). We should also mention that, for our waveguide configuration, this effect is generally much weaker than the non-uniformity issue.

### C. Effects from fabrication imperfections

We characterized the notch FWHM as 0.22 nm and the notch depth as 27.6 dB from Fig. 8(a), which is a zoomed-in view of Fig. 6(a). Both values are close to the target values of 0.20 nm and 30 dB, respectively. However, we notice that the notch shape errors could exist in experiment. In Fig. 8(b), the zoomed-in transmission of a 2 μm width spiral CWBG is shown, where the notch is shallower and wider than designed. This effect can also be seen by comparing Fig. 6(b) and Fig. 6(c), where the 2 μm CWBG spectrum has more noises.

These errors are found to vary from waveguide to waveguide, even with the same design. Therefore, we attribute them to fabrication imperfections. EBL current drift, EBL beam/stage movement error or etching rate non-uniformity would all cause this type of error. A wider waveguide design would increase the device tolerance to this effect, because the mode profile of a wider waveguide is less sensitive to width inaccuracy (Section III(B)).

Another concern of integrated photonics platform for wavelength-sensitive applications is wavelength shift due to fabrication imperfections. The wavelength inaccuracy for our device is generally < 100 pm, similar with our previous work [14]. Part of this inaccuracy is due to the systematic error in the fabrication, which results in an overall shift of all the notches. The thermal tuning technique could help to reduce this inaccuracy. For the $Si_3N_4/SiO_2$ platform, this technique is mostly based on the material's thermal-optic effect [36]. For our specific waveguide configuration, much of the optical mode resides in the $SiO_2$ cladding, which has a thermal-optic coefficient of about $10^{-5}$/K. We could derive that the required temperature variation range for a 100 pm shift would be only about 6 K.

## VI. CONCLUSION

To conclude, we have designed and characterized an on-chip arbitrary spiral filter with 55 uniform notches with insertion loss as low as 2.5 dB (throughput ∼ 56%). Notch FWHMs are about 0.22 nm and depths are about 28 dB. The whole device is 63 mm long, but with a footprint of only ∼ 1.5 mm thanks to the spiral design. The filter spectral range is also broadened to over 150 nm. Suffering from the additional loss caused by cladding mode coupling at shorter wavelengths, the overall throughput from 1450 to 1640 nm decreases to 35%. This is still close to the 39% throughput achieved in the most recent OH suppression filter fabricated with FBG [12]. Compared with the typical FBG notch depth 30 dB and notch FWHM 0.2 nm, our integrated device's performance is also on the same level. On the other hand, the integrated arbitrary filter we demonstrate here has a much more compact device size and potentially much lower mass-production cost. It could also provide stronger tuning capabilities when compared to optical fiber devices.

The arbitrary filter in this work is specially designed for astronomical observations, yet its successful demonstration also paves the road for many other applications from optical communications to quantum light control. It relies on layout patterning across a long distance with sub-10 nm resolution. Coherent grating interaction over a length of 50-mm is achieved with relatively small errors. This is a very positive sign for the R&D of large-scale photonic circuits with increasing flexibility and complexity.

## ACKNOWLEDGMENT

The authors acknowledge the support of the Maryland NanoCenter and its AIMLab.